\documentclass[twocolumn,aps,superscriptaddress]{revtex4}

\usepackage{graphicx}

\begin{document}
\pagestyle{empty} 
\title{Nanodroplets on rough hydrophilic and hydrophobic surfaces}

\author{C. Yang}
\affiliation{Institut f\"ur Festk\"orperforschung, Forschungszentrum J\"ulich,
D-52425 J\"ulich, Germany}

\author{U. Tartaglino}
\affiliation{Institut f\"ur Festk\"orperforschung, Forschungszentrum J\"ulich,
D-52425 J\"ulich, Germany}

\author{B.N.J. Persson}
\affiliation{Institut f\"ur Festk\"orperforschung, Forschungszentrum J\"ulich,
D-52425 J\"ulich, Germany}

\begin{abstract}
We present results of Molecular Dynamics (MD) calculations
on the behavior of liquid nanodroplets on rough hydrophobic and
hydrophilic solid surfaces.
On hydrophobic surfaces, the contact angle for nanodroplets depends strongly
on the root mean square roughness  amplitude, but it is nearly
independent of the fractal dimension of the surface.
Since increasing the fractal dimension increases the short-wavelength roughness,
while the long-wavelength roughness is almost unchanged, we conclude that for
hydrophobic interactions the short-wavelength (atomistic) roughness is not very important.
We show that the nanodroplet is in a Cassie-like state.
For rough hydrophobic surfaces, there is no contact angle hysteresis due to
strong thermal fluctuations, which occur at the liquid-solid interface 
on the nanoscale.
On hydrophilic surfaces, however, there is strong contact angle hysteresis 
due to higher energy barrier.
These findings may be very important for the development of artificially 
biomimetic superhydrophobic surfaces.
\vspace{1em}\\
{\it Reference:} Eur. Phys. J. E {\bf 25}, 139-152 (2008) \\
{\it DOI:} 10.1140/epje/i2007-10271-7 \\
{\it Preprint} arXiv:0710.3264

\end{abstract}
\maketitle


\section{Introduction}
\label{sec:intro}

In the year of 1805, Thomas Young and Pierre Simon de Laplace proposed that
an interface between two materials has specific energy, the so-called interfacial
energy, which is proportional to the interfacial surface area\cite{Young, Laplace, Rowlinson}.
This concept is the basis for
the field of wetting, which has become an extremely hot topic in the
last two decades\cite{Gennes,Quere}, thanks to biological and high-tech
applications, ranging from self-cleaning surfaces, microelectronics and thin film coatings, to image formation
that involve the spreading of liquids on solid surfaces.

Wetting describes the contact between a fluid and a solid surface.
Liquids with high surface tension (usually reflecting strong intra-molecular
bonds), or liquids on low-energy solid surfaces,
usually form nearly (complete) spherical droplets,
whereas liquids with low surface tension, or liquids on high-energy surfaces,
usually spread out on (or wet) the surfaces. This phenomenon is
a result of minimization of interfacial energy.
Thus, if a surface has a high free
energy, most liquids will spread on the surface since this will
usually lower the free energy.

Wetting phenomena have been widely studied both theoretically\cite{Chow,Patankar}
and experimentally\cite{Chen,Quere.soft} in connection with the physics of surfaces and interfaces.
The behavior of liquids on smooth solid surfaces is rather well
understood. However, for rough solid surfaces the situation is much less clear,
even though roughness occurs on practically
all real surfaces of engineering or biological interest.
Studies (and classification) of disordered and inhomogeneous
surfaces\cite{ca.hyst} should have significant impact on the problem of liquid
contact angle and wetting of rough
substrates\cite{droplet_prl,superwetting,confined.liquid,Ren,Bhushan}

\begin{figure}
  \begin{center}
  \includegraphics[width=0.35\textwidth]{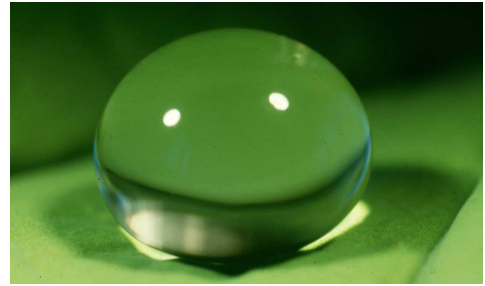}
  \end{center}
 \caption{ \label{exp_drop}
         A droplet on a superhydrophobic surface: The droplet touch
         the leaf only at a few points and forms a ball. It completely
         rolls off at the slightest declination.
         Adapted from Ref.\ \cite{link} with permission.}
\end{figure}

\begin{figure}
  \begin{center}
  \includegraphics[width=0.35\textwidth]{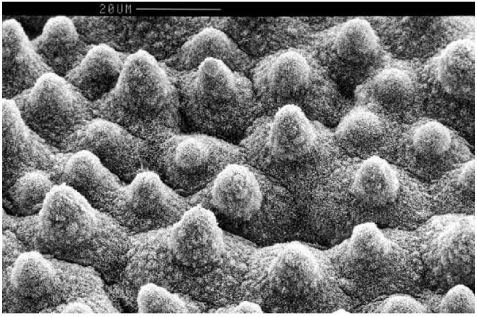}
  \end{center}
 \caption{ \label{Multiscale_rough}
         A leaf surface with roughness on several length scales optimized
         via nature selection for hydrophobicity and self-cleaning.
         Through the combination of microstructure (cells) and nanostructure
         (wax crystals) the macroscopic water contact angle $\theta_0$ is
         maximized. Adapted from Ref.\ \cite{link} with permission.}
\end{figure}

The fascinating water repellents of many biological surfaces, in particular
plant leaves, have recently attracted great interest for fundamental
research as well as practical
applications\cite{Planta,Botany,Dryplant,link,Kaogroup,naturematerial,Mimicking,
Lotusleaf}.
The ability of these surfaces to eliminate water beads completely
and thereby wash off contamination very effectively has been
termed the Lotus effect, although it is observed not only on the leaves of
the Lotus plant (Fig.~\ref{exp_drop}),
but also on many other plants such as strawberry, raspberry and
so on.
Water repellents are very important in many industrial and biological
processes, such as prevention of the adhesion of snow,
rain drops and fog to antennas,
self-cleaning windows and traffic indicators, low-friction surfaces and
cell mobility\cite{Nakajima, Coulson, Science1}.

Most leaves that exhibit strong hydrophobicity have
hierarchical surface roughness with micro- and nanostructures made of unwettable
wax crystals, which maximize the contact angle with water and most other
liquids. Fig.~\ref{Multiscale_rough} shows epidermal
cells (microscale roughness) covered with
wax crystals (nanoscale roughness). The wax crystals exhibit a 
relative high contact angle with water,
which is enhanced by the surface roughness.
Water droplets on the rough wax surface tend to
minimize the contact between the surface and the droplet by forming nearly
spherical droplets,
as approximately described by the two classical models due to Wenzel\cite{Wenzel} and
Cassie\cite{Cassie} (see below).
As a result the leaves have also a self-cleaning property:
because of the small adhesion energy (and small contact area) between
contamination particles and the rough leaf\cite{Planta}, during raining
water drops roll away removing the contamination particles from 
the leaf surface.

The hydrophobicity of solid surfaces is determined by both the chemical
composition and the geometrical micro- or nanostructure of the
surface\cite{Rul,Dup,Chen}.
Understanding the wetting of corrugated and porous surfaces
is a problem of long standing interest in areas ranging from textile
science\cite{Textile} to catalytic reaction engineering\cite{Catalytic}.
Renewed interest in this problem has been
generated by the discoveries of surfaces with small scale
corrugations that exhibit very large contact angles for water and other
liquids---in some cases the contact angle is close to $180^{\circ}$. Such surfaces
 are
referred to as superhydrophobic\cite{Pearldrops}.

In this paper we present results of Molecular Dynamics (MD) calculations
on the behavior of liquid nanodroplets on rough hydrophilic and
hydrophobic solid surfaces. We find that for
hydrophobic surfaces, the contact angle for nanodroplets depends strongly
on the root mean square surface roughness amplitude, but is nearly
independent of the fractal dimension $D_{\rm f}$ of the surface.
For hydrophobic rough surfaces we do not detect any contact angle hysteresis.
Both results can be explained by the strong
thermal fluctuations which occur at the liquid-solid interface on the nanoscale.
On hydrophilic surfaces, however, strong contact angle hysteresis has been found
 due to
the higher energy barrier for interfacial liquid density fluctuations.
These findings may be crucial for the development of artificial biomimetic superhydrophobic surfaces.

\begin{figure}
  \begin{center}
  \includegraphics[width=0.35\textwidth]{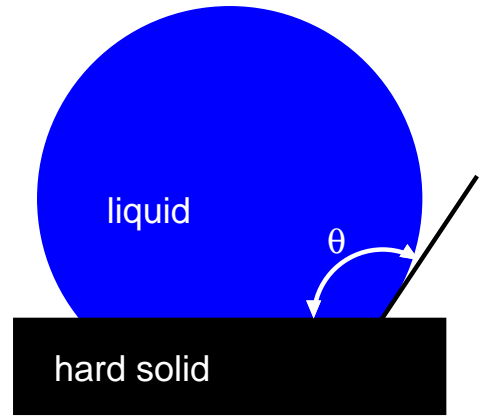}
  \end{center}
    \caption{ \label{contactangle}
     Liquid droplet on flat substrate. The contact angle $\theta$ is between
     $0$ (complete wetting) and $\pi$.}
\end{figure}

\section{Theoretical background} 
\label{sec:theory}

In this section we briefly describe some results from the theory of the liquid-solid contact
angle, which are necessary for the interpretation of the numerical results presented
in Sec.~\ref{sec:numerical}. We emphasize the importance of thermal
fluctuations for the contact dynamics at the
nanoscale as compared to micrometer or macroscopic dimensions. 

\subsection{Flat surfaces}
\label{sec:flatsurfaces}

If gravitational effects can be neglected, a liquid droplet on a flat substrate
forms a spherical cap, see Fig.~\ref{contactangle}.
The contact angle $\theta$
is determined by the minimization of the free energy and depends on
the interfacial free energies per unit area: solid/liquid $\gamma_{\rm sl}$,
solid/vapor $\gamma_{\rm sv}$ and liquid/vapor $\gamma_{\rm lv}$.
Minimizing of the surface free energy, with the constrain of fixed
volume of the droplet, gives the Young's equation, first proposed
by Thomas Young about two hundred years ago:

\begin{equation}
  \label{Young_equation}
  \gamma_{\rm sl}+\gamma_{\rm lv}{\rm cos} \theta =\gamma_{\rm sv}
\end{equation}

Complete wetting corresponds to $\theta=0$, and typically
happens for liquids with low surface tension
$\gamma_{\rm lv}$, and on
solids with high surface energy
$\gamma_{\rm sv}$.
Liquids with high surface tension on surfaces with low surface energy
tend to form droplets with
high contact angle $\theta$.
Eq. (\ref{Young_equation})  was deduced for a substrate which is assumed to be perfectly smooth, homogeneous, and rigid.
However, in reality, structured or rough surfaces are quite common. So it's necessary to know how the contact angle behaves on rough surfaces.

\begin{figure}
  \begin{center}
  \includegraphics[width=0.45\textwidth]{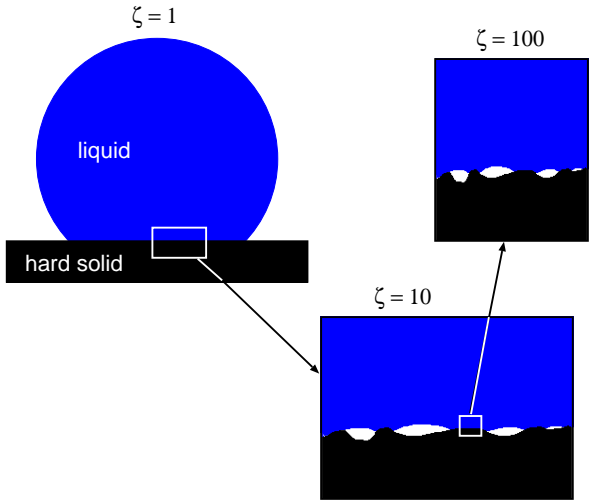}
  \end{center}
    \caption{ \label{liquid.drop.magnification}
     Liquid droplet on a rough substrate. At the lowest magnification $\zeta$ 
     the surface appears flat and the liquid contact angle is $\theta_0$. 
     At increasing magnification surface roughness is observed and the liquid 
     will in general only make contact with the substrate in some asperity 
     contact regions.}
\end{figure}

\subsection{Rough surfaces: minimum free energy state}
\label{sec:roughsurfaces1}

Most surfaces of practical interest have roughness on many different length 
scales. For simple periodic surface profiles one may develop accurate 
analytical treatments of the liquid droplet contact angle 
(see e.g., Ref.\ \cite{Carbone}), but for randomly rough surfaces
the situation is much more complex. For surfaces with random roughness, e.g., 
self-affine fractal surfaces (see below),
one may develop a general theory based on the study of
the system at different magnifications $\zeta$, 
see Fig.~\ref{liquid.drop.magnification}.
Here $\zeta= D/\lambda$ where $D$ is the diameter of the
droplet-substrate (apparent) contact area and $\lambda$ the resolution. One can introduce effective
interfacial liquid-solid and solid-vapor free energies
(per unit area) $\gamma_{\rm sl} (\zeta)$ and $\gamma_{\rm sv} (\zeta)$
which depend on the magnification $\zeta$\cite{cappilary}. At the highest magnification 
$\zeta_1$, corresponding to nanometer (or atomistic) resolution, 
these quantities reduce to those for the flat surface\cite{comment1},
$\gamma_{\rm sl}(\zeta_1) = \gamma_{\rm sl}$ and $\gamma_{\rm sv}(\zeta_1)=\gamma_{\rm sv}$.
Since the substrate appears flat at the lowest magnification $\zeta=1$, 
the macroscopic contact angle (corresponding to $\zeta=1$) is obtained using 
the Young's equation with $\gamma_{\rm sl}$ and $\gamma_{\rm sv}$ replaced by
$\gamma_{\rm sl}(1)$ and $\gamma_{\rm sv}(1)$, i.e.

\begin{equation}
  \label{Young_equation_1}
  \gamma_{\rm sl}(1)+\gamma_{\rm lv}\cos\theta_0 =\gamma_{\rm sv}(1)
\end{equation}
The change in the surface free energy (per unit area)
when a liquid with a flat surface is brought in contact with the
substrate is
$$\Delta F /A_0 = \gamma_{\rm sl}(1)-\gamma_{\rm sv}(1)-\gamma_{\rm lv} = 
-\gamma_{\rm lv} (1+\cos\theta_0)$$
where $A_0$ is the (projected) surface area. Note that increasing 
contact angle $\theta_0$ corresponds to a increasing interfacial free energy. Thus,
if a liquid drop can occur in several metastable states on a surface, the state
with the smallest contact angle corresponds to the (stable) 
minimal free-energy state.

Using Eq.\ (\ref{Young_equation_1}) it is trivial to derive the results of the
so called 
Wenzel\cite{Wenzel} and Cassie\cite{Cassie} models.
In the Wenzel model it is assumed that complete contact occurs at the 
liquid-solid interface. Thus
\begin{equation}
  \label{bbluu}
\gamma_{\rm sv}(1)=r \gamma_{\rm sv}(\zeta_1) \,, \qquad
\gamma_{\rm sl}(1)=r\gamma_{\rm sl}(\zeta_1) \,,
\end{equation}
where $r=A/A_0>0$ is the ratio between the surface area $A$ of the rough 
substrate, and the projected (or nominal) surface area $A_0$.
Substituting (\ref{bbluu}) into (\ref{Young_equation_1}) gives the contact
angle $\theta_0$ on the rough 
surface in terms of the contact angle $\theta$ on the microscopically flat 
surface of the same material (Wenzel equation):
\begin{equation}
  \label{Wenzel_equation}
  \cos \theta_0 = r \cos \theta \,.
\end{equation}

\begin{figure}
  \begin{center}
  \includegraphics[width=0.45\textwidth]{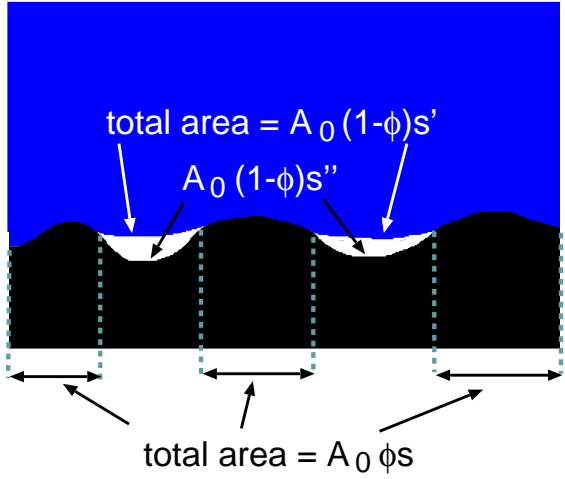}
  \end{center}
    \caption{ \label{scem.fig}
     The interface between liquid and solid. $\phi < 1$ is the ratio between 
     the projected liquid-solid contact area and the nominal (or apparent)
     contact area $A_0$.
     Because the solid surface is curved, the actual liquid-solid contact area 
     will be $A_0 \phi s$ where $s>1$. Similarly, since in general the 
     liquid-vapor interface is curved (in spite of the fact that the total 
     curvature $1/R_1+1/R_2$ may vanish) and tilted (relative to the average
     surface plane) the total liquid-vapor interface area
     is $A_0 (1-\phi)s'$, with $s'>1$. Similarly, the solid-vapor interface
     area equals $A_0 (1-\phi)s''$ with $s'' >1$.}
\end{figure}

In the Cassie model\cite{Cassie} it is assumed that some air (or vapor)
remains trapped between the drop and the cavities of the rough surface. 
In this case the interface free energy
\begin{equation}
   \label{blii1}
\gamma_{\rm sv}(1)=r \gamma_{\rm sv}(\zeta_1),
\end{equation}
\begin{equation}
   \label{blii2}
\gamma_{\rm sl} (1) = \phi r \gamma_{\rm sl}(\zeta_1)+(1-\phi)(r \gamma_{\rm sv}(\zeta_1)+\gamma_{\rm lv}),
\end{equation}
where $\phi$ is the fraction of the (projected) area where the liquid is in contact 
with the solid. Substituting (\ref{blii1}) and (\ref{blii2}) in
(\ref{Young_equation_1}) gives
\begin{equation}
   \label{Cassie_equation}
  \cos \theta_0 = r \cos \theta - (1-\phi) (1+r \cos \theta) \,.
\end{equation}
Note that for $\phi = 1$, (\ref{Cassie_equation}) reduces to
(\ref{Wenzel_equation}). In the original Cassie model it was assumed that
$r=1$.
We note that while the Wenzel theory is exact if the liquid is in contact with
the substrate everywhere within the nominal liquid-substrate contact area,
the Cassie theory is always approximate and often
not very accurate. This is easily understood from Fig.~\ref{scem.fig} 
which shows the interface between a liquid and a solid. $\phi < 1$ is the ratio
between the projected liquid-solid contact area and the nominal (or 
apparent) contact area $A_0$.
Because the solid surface is curved, the actual liquid-solid contact area will
be $A_0 \phi s$ where $s>1$. Analogously, since in general the liquid-vapor 
interface is curved (in spite of the fact that the total curvature 
$1/R_1+1/R_2$ may vanish)
and tilted (relative to the average surface plane), the total liquid-vapor 
interface area is $A_0 (1-\phi)s'$, with $s'>1$. Similarly, the solid-vapor 
interface area equals $A_0 (1-\phi)s''$ with $s'' >1$. In deriving
(\ref{Cassie_equation}) it is assumed that $s=s''=r$ and $s'=1$. 

Of the two states, Cassie and Wenzel, the stable one, that is the one with
lower free energy, is the one with larger $\cos\theta_0$.
Comparing (\ref{Wenzel_equation}) and (\ref{Cassie_equation}) shows that the value of $\cos\theta_0$ for the Cassie
state is larger if
$1+r \cos \theta < 0$ or
\begin{equation}
  \label{coss}
  \cos \theta < -1/r \,.
\end{equation}
Since $r$ is a measure of the magnitude of the surface roughness, we may qualitatively state
that only for hydrophobic surfaces (with $\cos \theta < 0$ or $\theta > 90^{\circ}$) with large enough
roughness (i.e., large enough $r=A/A_0$) will the Cassie state be the thermodynamically stable state.

The approach described above, where the interface is studied at different 
magnifications, is very general and a similar approach has recently been 
developed for the contact mechanics between elastic solids with randomly 
rough surfaces\cite{Persson2002} (see also 
Ref.\ \cite{Herminghaus}).

It is well known that the roughness of a hydrophobic solid 
(with $\theta > 90^{\circ}$ on the flat substrate) enhances its hydrophobicity.
If the contact angle of water on such flat solids is of the order
of $100^{\circ}$ to $120^{\circ}$, on a rough or microtextured surface it may be
as high as $150^{\circ}$ to $175^{\circ}$\cite{Herminghaus,Superstates,Science1}.
Both the Wenzel model and the Cassie model can explain this effect.

\begin{figure}
  \begin{center}
  \includegraphics[width=0.30\textwidth]{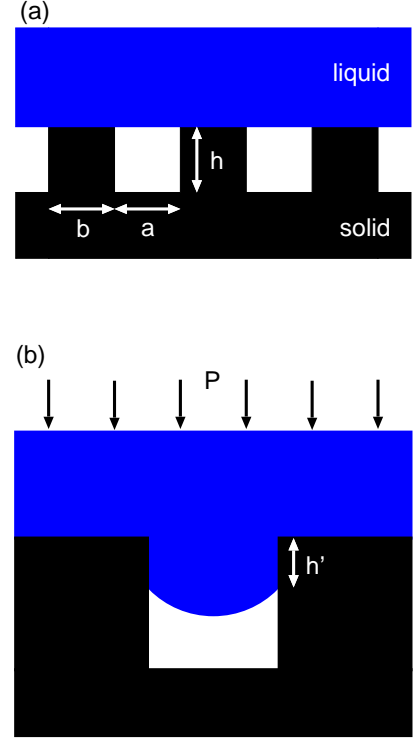}
  \end{center}
    \caption{ \label{activation}
     (a) Liquid drop (in the Cassie state) in contact with a surface with periodic surface roughness.
     (b) Even if the Cassie state (incomplete liquid-solid contact) is the ground state, with an applied
     pressure $p$ one can squeeze the droplet into the Wenzel state.}
\end{figure}

\begin{figure}
  \begin{center}
  \includegraphics[width=0.4\textwidth]{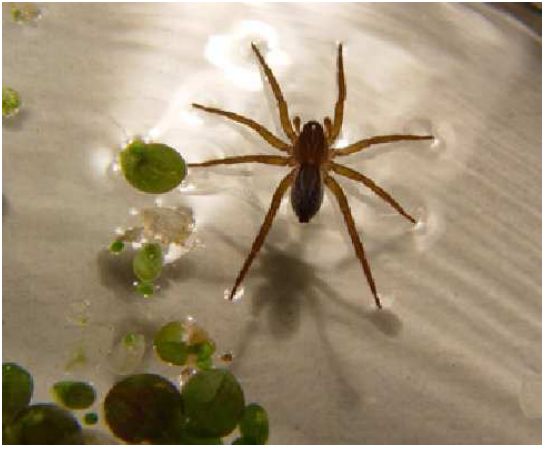}
  \end{center}
    \caption{ \label{spider}
     Water spiders have thin hydrophobic (wax coated) hair with nanoscale roughness
which trap air and enhance hydrophobicity.  
}
\end{figure}

Let us consider the simplest surface roughness consisting of a periodic rectangular
roughness profile as illustrated in Fig.~\ref{activation} (a) ($xz$-plane). The free energy 
(per unit surface area) for the
Cassie state shown in the figure is
$$\gamma_{\rm C} = \left [(a+2h)\gamma_{\rm sv}+a\gamma_{\rm lv}+b\gamma_{\rm sl}\right ]/(a+b)$$
The free energy for the Wenzel state (complete contact) is
$$\gamma_{\rm W} = (a+b+2h)\gamma_{\rm sl}/(a+b)$$
Using (\ref{Young_equation}) we can write the difference in free energy
$$\gamma_{\rm C}-\gamma_{\rm W} = \gamma_{\rm lv} [a(1+{\rm cos} \ \theta )+2h{\rm cos} \ \theta ]/(a+b)$$ 
Thus, the Cassie state has a lower free energy than the Wenzel state if
\begin{equation}
  \label{condition}
   {\rm cos} \ \theta < - \left (1+{2h\over a}\right )^{-1}
\end{equation}
which is satisfied only if for the flat surface
$\theta > 90 ^{\circ}$, and if the ratio $h/a$ is large enough. 
Note that in this case $r=A/A_0 = (a+b+2h)/(a+b)$ so the (approximate) criteria (\ref{coss}) reduces to
$${\rm cos} \ \theta < - \left (1+{2h\over a+b}\right )^{-1}$$
which is of similar general form as (\ref{condition}).
In Nature strongly
hydrophobic surfaces are often obtained by covering the surface with thin, long
(so that $h/a \gg 1$) hydrophobic fibers. Thus, insects which move on top of water, e.g., water spiders
(see Fig.~\ref{spider})
have a high density of thin wax coated hair on their legs. In addition, the hair fibers have nanoscale
roughness which traps air and enhances the hydrophobicity\cite{WaterS}. 
In this case the water-leg
contact will be in the Cassie state even when the insect is squeezed towards the water by the
weight of the insect. 

\subsection{Rough surfaces: activation barriers and hysteresis}
\label{sec:roughsurfaces2}

Consider a cylindrical cavity as in Fig.~\ref{activation} (b) and assume first
the Cassie state as in the figure. Let us apply a pressure $p$ to the droplet. In this case the
liquid will bend inwards in the cavity and if the applied pressure is larger than a critical
value $p_c$, the liquid will
be squeezed into the cavity (we assume that the air in the cavity can leave the cavity, e.g., diffuse into
the liquid). It is easy to show that the pressure 
\begin{equation}
  \label{pressure}
  p_c = -2\gamma_{\rm lv} {\rm cos} \theta /R
\end{equation}
where $R$ is the radius of
the cavity. 
To prove this relation, note that the pressure work to squeeze the liquid a distance $h'$ into the cavity
(see Fig.~\ref{activation}) is given by $p_c \pi R^2 h'$ and this must equal the change in interfacial
free energy which equals $2\pi R h' (\gamma_{\rm sl}-\gamma_{\rm sv})$. Using these equations and (\ref{Young_equation})
gives (\ref{pressure}).

From (\ref{pressure}) it follows that if 
$\theta < 90^{\circ}$ (hydrophilic interaction), $p_c < 0$ and the liquid will be spontaneously
sucked into the cavity and will fill out the cavity. If $\theta > 90^{\circ}$ 
(hydrophobic interaction), $p_c > 0$ and   
for nanometer sized cavities, the pressure $p_c \sim 100 \ {\rm MPa}$, so very high pressures are
necessary for squeezing the liquid into narrow cavities. However, if the liquid is squeezed into the cavity
and completely fills the cavity, then the resulting Wenzel state is (at least) metastable. However, 
for nanometer sized cavities thermal fluctuations may give rise
to strong local fluctuations between the Cassie (empty cavity) and Wenzel (filled cavity) states.
This is easy to understand since the 
energetic barrier (for a hydrophobic system)
for going from the Cassie state to the Wenzel state
will be of order $\varepsilon \sim p_c \pi R^2h = - 2\pi R h \gamma_{\rm lv}{\rm cos}\theta $,
and strong fluctuations on macroscopic time scales will occur as long as $\varepsilon \approx
0.7 \ {\rm eV}$ or less, and strong fluctuations on the nanosecond time scale occur if
$\varepsilon \approx 0.4 \ {\rm eV}$ or less [note: the rate to jump over a barrier of height $\varepsilon$ is
$w=\nu\,\exp(-\varepsilon /k_{\rm B}T)$ where typically $\nu \approx 10^{12} \ {\rm s}^{-1}$; 
at room temperature $ w \approx 1 \ {\rm s}^{-1}$ if $\varepsilon \approx 0.7 \ {\rm eV}$ and
$ w \approx 10^9 \ {\rm s}^{-1}$ if $\varepsilon \approx 0.4 \ {\rm eV}$].
In a typical case this condition is satisfied as long as 
$R$ and $h$ are of order of one
nanometer or less. In our computer simulations we do indeed observe very strong thermal
fluctuations at the liquid-solid interface, in particular for rough hydrophobic surfaces, see
Sec.~\ref{sec:hydrophobic}.

The Wenzel droplets are highly pinned, 
and the transition from the Cassie to the Wenzel state results in the loss of 
the anti-adhesive properties generally associated with superhydrophobicity.
However, for nanodroplets on rough hydrophobic surfaces, we find that the
Wenzel state is unstable: if the droplet is pressed into complete contact with the 
substrate (Wenzel-like state) and then let free,
it quickly jumps back to the Cassie-like state due to strong thermal 
fluctuations. For a macroscopic droplet on surfaces with long wavelength
roughness, the energetic barrier towards flipping 
from the Wenzel to the Cassie state may be so large that, even if the Cassie 
state is the minimum free energy configuration, the system may remains trapped in 
the (metastable) Wenzel state for all time periods of physical relevance.

\begin{figure}
  \begin{center}
  \includegraphics[width=0.45\textwidth]{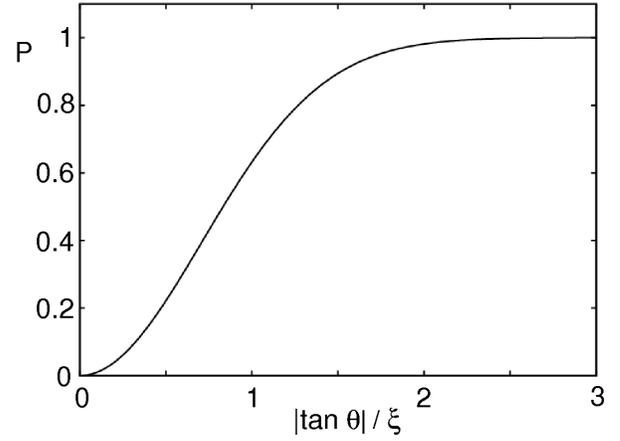}
  \end{center}
 \caption{ \label{Ptantxi}
          The fraction $P$ of the surface area where the absolute value of 
          the slope is smaller than $|{\rm tan \theta}|$
          as a function of $|{\rm tan \theta}| / \xi$.
          See text for details.}
\end{figure}

\subsection{Cassie and Wenzel states for randomly rough surfaces}
\label{sec:roughsurfaces4}

In this section we discuss under which condition one expects the Cassie state or the Wenzel state
to prevail. Consider a rough surface and let $z=h({\bf x})$
be the height of the surface at the point ${\bf x}=(x,y)$.
A randomly rough surface can be obtained by adding plane waves with random phases:
$$h({\bf x}) = \sum_{\bf q} B({\bf q})e^{i[{\bf q}\cdot {\bf x}+\phi({\bf q})]}$$
where $\phi({\bf q})$ are independent random variables, uniformly distributed in the 
interval $[0,2\pi [$, and with $B({\bf q})= (2\pi /L)[C({\bf q})]^{1/2}$, where $L=A_0^{1/2}$
is the linear size of the surface. The surface roughness power spectrum: 
$$C({\bf q})=\frac{1}{(2\pi)^2} \int d^2x \ \langle h({\bf x}) h({\bf 0}) 
\rangle  e^{i {\bf q\cdot x}}.\eqno(11)$$
Here $h({\bf x})$ is the surface height profile and $\langle\cdots\rangle$
stands for ensemble average. We have assumed that $\langle h({\bf x})\rangle = 0$.
We assume that the statistical properties of the rough surface are isotropic, so that $C({\bf q})$
only depends on the magnitude $q=|{\bf q}|$ of the wave vector ${\bf q}$.

For randomly rough surfaces the normalized surface area
$r=A/A_0$ is given by (see Appendix):
$$r = \int_0^\infty dx \ \left (1+x\xi^2\right )^{1/2}e^{-x}\eqno(12)$$
where
$$\xi^2 = \int d^2q \ q^2C(q) = 2 \pi 
\int_0^\infty dq \ q^3C(q) = \langle (\nabla h )^2\rangle\eqno(13)
$$
is the square of the average slope. 

The fraction of the surface where the surface slope $s < s_0$ is given by (see Appendix):
$$P(s_0)=1-e^{-(s_0/\xi)^2}.$$
Note that as $\xi \rightarrow 0$ (corresponding to a flat surface) $P(s_0)\rightarrow 1$
which is expected because the slope of a flat surface is zero and hence smaller than
any finite value $s_0$. Assume that a liquid exhibits the contact angle $\theta$ on the
perfectly flat substrate. The fraction of the surface where the slope 
$| \nabla h({\bf x}) | < |{\rm tan} \theta|$ is given by 
$$P({\rm tan} \theta )=1-e^{-({\rm tan} \theta/\xi)^2}.\eqno(14)$$
This function is shown in Fig.~\ref{Ptantxi}. If we assume that the liquid surface in the 
liquid-solid non-contact
region is flat and parallel to the average substrate surface plane, then we expect the 
liquid to only occupy
the region where the slope is smaller than ${\rm tan}\theta$.
Note that more than $90\%$ of the surface area will 
have a slope below $|{\rm tan} \theta|$ if $|{\rm tan} \theta| / \xi > 1.5$ and in this case the 
Wenzel state will tend to prevail, while more than $90\%$ of the surface will have  a 
slope above $|{\rm tan} \theta|$ if $ |{\rm tan} \theta| / \xi < 0.3$ and in this
case the Cassie state will tend to prevail. For the system we study 
below $\xi < 2$ (see Fig.~\ref{ca_ratio})
and for the hydrophobic system $\theta \approx 103^{\circ}$ we get $|{\rm tan}\theta| / \xi > 2.2$.
Thus, one would expect the Wenzel state to prevail. However, the numerical data (see below) tend
to suggest that the system is in a Cassie-like state. We attribute this to the strong fluctuations
at the liquid-solid contact which occurs at the nanoscale, which are particularly important for
nanoscale droplets. 

\begin{figure}
  \begin{center}
  \includegraphics[width=0.5\textwidth]{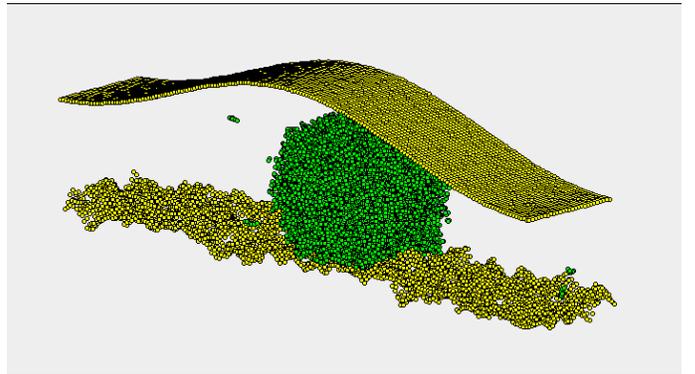}
  \end{center}
 \caption{ \label{sideview_snap}
           3D side-view snapshot of a octane liquid droplet on a hydrophobic and rough
           substrate.
           The rigid substrate comprises $200\times30$ atoms disposed on a square lattice
           with lattice constant $a=2.53$ \AA. These atoms have been randomly displaced along
           the $z$ coordinate, orthogonal to the wall, so to reproduce the desired roughness.
           The Lennard-Jones solid-liquid interaction potential 
           $V(r)=4\epsilon[(r_{0}/r)^{12}-(r_{0}/r)^{6}]$ with
           $r_{0}=3.28$ \AA, $\epsilon=4$ meV for hydrophobic substrate and
           $\epsilon=8 \ {\rm meV}$ for hydrophilic substrate.} 
\end{figure}

\section{Simulation Method}
\label{sec:simulation}

We have used Molecular Dynamics (MD) to study the contact angle and contact angle hysteresis.
Here we briefly describe the system we studied and how we generated the rough substrate 
surfaces.

\subsection{Molecular dynamics model}
\label{sec:moldyn}

\begin{figure}
  \begin{center}
  \includegraphics[width=0.4\textwidth]{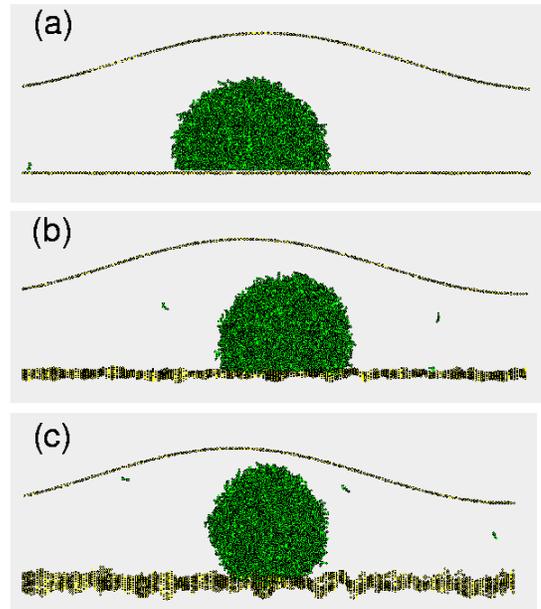}
  \end{center}
 \caption{ \label{snapshots_diff_rms}
           Snapshots for different root mean square roughness. (a) the droplet
           is in contact with the flat substrate. (b) and (c) are for rough 
           substrates with the root mean square amplitude $\sigma =2.3$ \AA\ 
           and $\sigma =4.8$ \AA, respectively. Adapted from Ref.\ \cite{droplet_prl}.}
\end{figure}

We have used MD calculations to study the
influence of surface roughness on liquid droplet contact angle 
and contact angle hysteresis.
We have studied hydrocarbon liquid droplets on different self-affine fractal
surfaces. The nanodroplets contained 2364 octane molecules
$\rm{C}_8\rm{H}_{18}$ at $T= 300 \ {\rm K}$, which is between the melting
and boiling points of octane.
The fractal surfaces were generated by adding plane waves with random phases (see Sec.~IID
and Ref.\ \cite{Yang}).
Periodic boundary conditions are applied along the $x$ and $y$ directions.
The periodically repeated cell forms a rectangle $L_x \times L_y$ with
$L_x=506$ \AA\ and $L_y=75.9$ \AA\ (see Fig.~\ref{sideview_snap}).
The (non-contact) cylindrical droplet diameter is about $104$ \AA, and the size of the 
droplet-substrate contact area varies (for the hydrophobic system) 
from $\approx 115$ \AA\ (case (a)
in Fig.~\ref{snapshots_diff_rms}) to $\approx 60$ \AA\ (case (c)).

For most real surfaces usually there is a roll-off wave vector $q_0$,
below which the power spectrum $C(q)$ of the surface is approximately constant. For
$q > q_0$ we assume the power spectrum has the power-law behavior $C(q) \sim
q^{-2(H+1)}$\cite{Yang} corresponding to a self affine fractal surface with the
fractal dimension
$D_{\rm f} = 3-H$.
Different fractal surfaces are obtained by changing the root mean square (rms)
roughness amplitude $\sigma$, and the fractal dimension $D_{\rm f}$.
The roll-off wave-vector for the rough surface is $q_0=2\pi/L$ with $L=38$ \AA,
and the magnitude of the short-distance cut-off wave vector $q_1=\pi/a$, 
where $a= 2.53$ \AA\ is the substrate lattice constant. In the present work,
we are mainly interested in how the rough structure of the substrate influence wetting.
A curved upper wall has been used to speed up the formation of the droplet in the 
initial preparation\cite{Vladimir} and to limit the gas-phase 
volume so that the droplet cannot (fully) evaporate. 
In some simulations we used a flat upper wall in order to be able to push the droplet 
towards the substrate a bit more, in order to study the evolution of the receding contact angle. 

The lubricant molecules are described through the Optimized Potential
for Liquid Simulation (OPLS)\cite{jorgensen1984x1,dysthe2000x1}; this
potential is known to provide density and viscosity of hydrocarbons
close to the experimental one.
Each octane molecules comprises four units (particles), each particle corresponding to
one chemical group $\rm CH_2$, $\rm CH_3$. The interaction between particles of different
molecules is described by Lennard-Jones potentials. The intramolecular interactions
include two body forces that keep the bond length $\rm C-C$ close to 1.53 \AA,
three body forces imposing a preferred angle of $115^{\circ}$ between the carbon atoms,
and four body forces favoring a well defined torsion of the molecules. The four body 
forces apply to the sequence of carbon atoms $\rm C-C-C-C$\cite{Ugo}. 

\begin{figure}
  \begin{center}
  \includegraphics[width=0.35\textwidth]{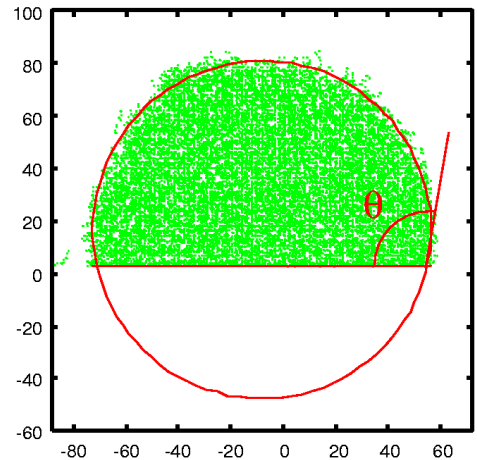}
  \end{center}
 \caption{ \label{contact_angle_flat}
           Determination of the contact angle $\theta$ for the flat
           substrate. Side view. Adapted from Ref.\ \cite{droplet_prl}.}
\end{figure}

We used the Lennard-Jones interaction potential between
droplet atoms and substrate atoms.
The L.-J. parameters for a hydrophobic surface are chosen such
that the Young contact angle is about $100^{\circ}$ when a droplet sits on
the flat surface.
Because of the periodic boundary conditions and the size of our system,
the liquid droplet forms a cylinder with the central line
along the $y$-axis, as shown in Figures \ref{sideview_snap} and \ref{snapshots_diff_rms}.
We fit the density profile of the droplet to a cylinder (see Fig.~\ref{contact_angle_flat}),
and obtain the contact
angle $\theta=103^{\circ}$ 
for the droplet in contact with the flat hydrophobic substrate, while
for the flat hydrophilic substrate $\theta=39\pm3^{\circ}$.

\begin{figure}
  \begin{center}
  \includegraphics[width=0.45\textwidth]{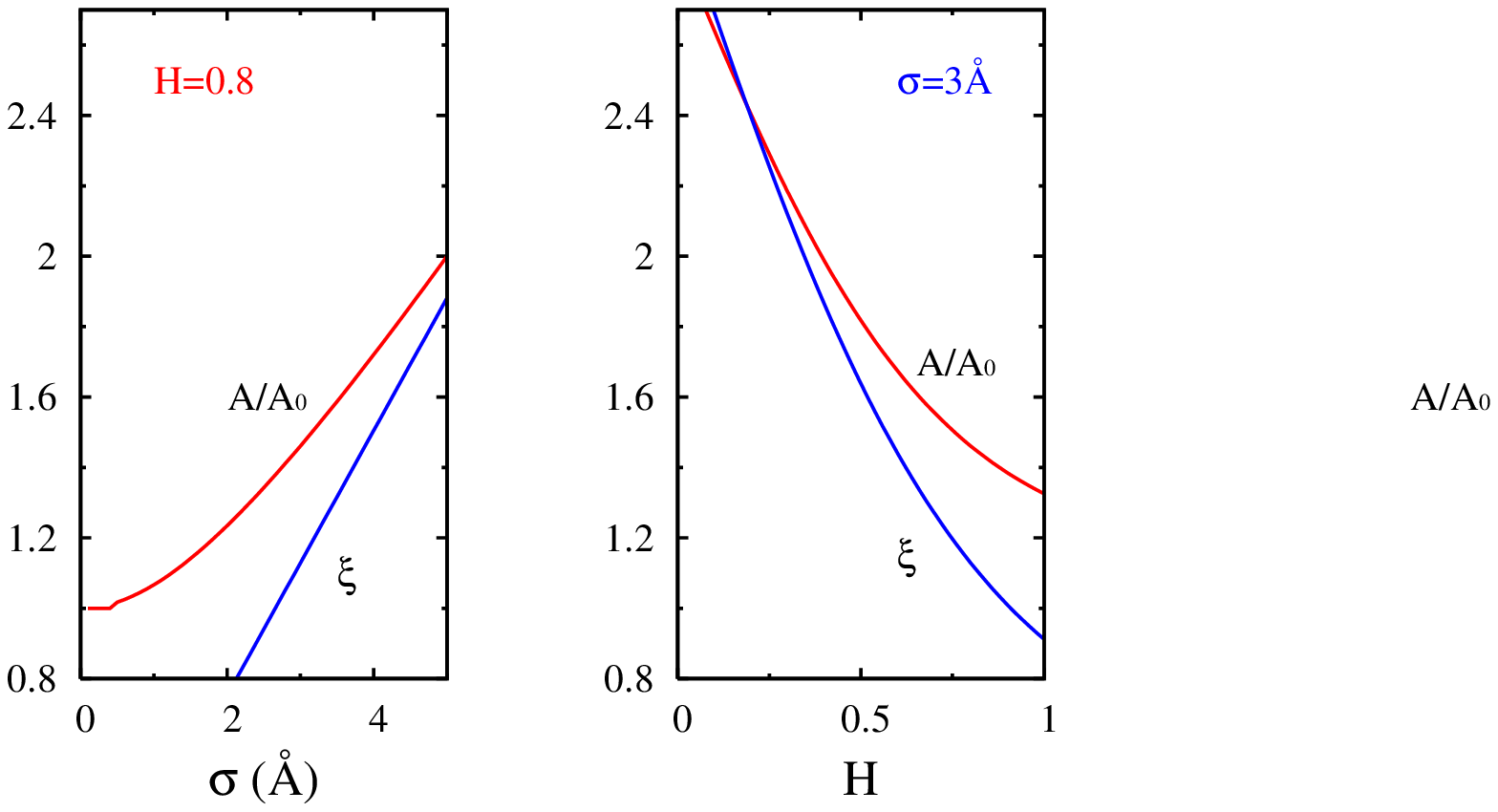}
  \end{center}
 \caption{ \label{ca_ratio}
           The average slope $\xi$ and the ratio $A/A_0$ between the actual $A$ and the nominal 
           (or projected) $A_0$ surface area, as a function of the root mean 
           square roughness $\sigma$ when Hurst exponent $H=0.8$, and as a
           function of Hurst exponent $H$ for $\sigma =3$ \AA.}
\end{figure}

\subsection{Multiscale rough surfaces}
\label{sec:multiscale}

Many solid surfaces in nature, e.g., surfaces prepared by fracture (involving crack propagation),
tend to be nearly self-affine fractal.
Self-affine fractal surfaces have multiscale roughness, sometimes extending from
the lateral size of the surface down to the atomic scale.
A self-affine fractal surface has the property
that if part of the surface
is magnified, with a magnification which in general is appropriately
different in the directory perpendicular to the surface as compared to
the lateral directions, the surface ``looks the same''\cite{P3}
i.e., the statistical
properties of the surface are invariant under this scale transformation.

The most important property of a randomly rough surface is the surface
roughness power spectrum $C(q)$.
We assume that the statistical properties of the surface are
translational invariant and isotropic so that $C(q)$ depends only on the
magnitude $q=|{\bf q}|$ of the wavevector ${\bf q}$.
For a self-affine surface the power spectrum has the
power-law behavior $C(q) \sim  q^{-2(H+1)}$, where the Hurst exponent
$H$ is related to the fractal dimension $D_{\rm f}=3-H$.
Of course, for real surfaces
this relation only holds in some finite wave vector region
$ q_{0} < q < q_{1}$.
Note that in many cases there is roll-off wavevector $q_{0}$ below which
$C(q)$ is approximately constant. We have generated self affine fractal surfaces by
adding plane waves with random phases and appropriately chosen weights, as
described in detail in Ref.\ \cite{P3, Yang}.

In Fig.~\ref{ca_ratio} we show the average slope $\xi$ and the ratio $A/A_0$ between the surface
area $A$ and the nominal 
(or projected) $A_0$ surface area, as a function of the root mean 
square roughness $\sigma$ when Hurst exponent $H=0.8$, and as a
function of Hurst exponent $H$ for $\sigma =3$ \AA.

\begin{figure}
  \begin{center}
  \includegraphics[width=0.45\textwidth]{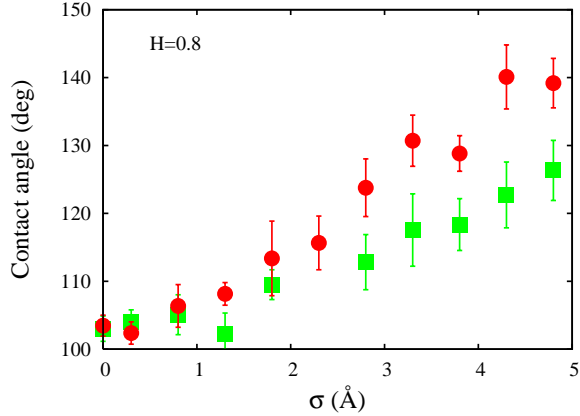}
  \end{center}
 \caption{ \label{ca_vs_rms}
          The contact angle as a function of the root mean square roughness 
          $\sigma$. The circle points are numerical results from the 
          simulations, while the square points are obtained from the Cassie
          model (see Eq.~\protect\ref{Cassie_equation}).
          Each data point is an average over several snap-shot configurations.
          The fractal dimension is $D_{\rm f}=2.2$. Adapted from Ref.\ \cite{droplet_prl}.}
\end{figure}

\begin{figure}
  \begin{center}
  \includegraphics[width=0.45\textwidth]{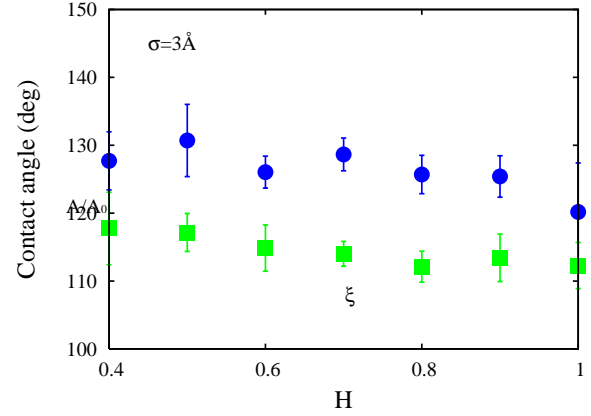}
  \end{center}
 \caption{ \label{ca_vs_H}
           The contact angle $\theta$ as a function of  Hurst exponent $H$
         for the rms roughness $\sigma =3$ \AA.
         The circles and squares have the same meaning as that in
         Fig.~\protect\ref{ca_vs_rms}
         The fractal dimension is $D_{\rm f}=3-H$. Adapted from Ref.\ \cite{droplet_prl}.}
\end{figure}

\begin{figure}
  \begin{center}
  \includegraphics[width=0.45\textwidth]{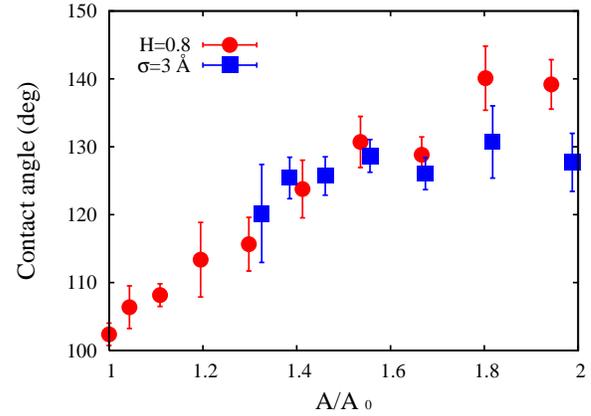}
  \end{center}
 \caption{ \label{rms.H.ca.vs.A}
           The contact angle $\theta$ as a function of the normalized
          surface area $A/A_0$ when the root-mean-square amplitude $\sigma$ increases
         for fixed Hurst exponent ($\rm H=0.8$, the solid circles) (from Fig.~\ref{ca_vs_rms}) and when the Hurst exponent decreases
         for a fixed root-mean-square amplitude ($\sigma = 3$ \AA, the solid squares) (from Fig.~\ref{ca_vs_H}).} 
\end{figure}

\section{Numerical results}
\label{sec:numerical}

We present numerical results for the contact angle and contact angle hysteresis
for both hydrophilic and hydrophobic systems. The substrate surfaces are assumed
to be self affine fractal, but we have varied the fractal dimension and the 
root-mean-square roughness amplitude.

\subsection{Static contact angle on hydrophobic surface}
\label{sec:staticangle}

The apparent contact angle, $\theta_0$, as a function of the root mean square 
roughness (rms), is shown in Fig.~\ref{ca_vs_rms} with the fractal dimension 
$D_{\rm f}=2.2$. There is a strong increase in $\theta_0$ with
increasing rms-roughness amplitude. Fig.~\ref{ca_vs_H} shows how $\theta_0$
depends on the Hurst exponent $H=3-D_{\rm f}$. Note that $\theta_0$ is almost
independent of $H$.

\begin{figure}
  \begin{center}
  \includegraphics[width=0.45\textwidth]{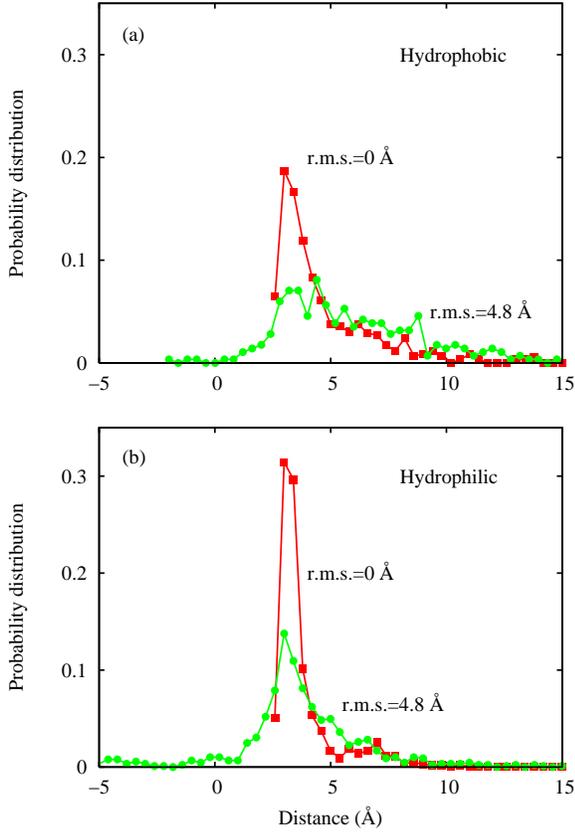}
  \end{center}
    \caption{ \label{histogram_droplet}
           (a) The height probability distribution for hydrophobic surface
           both flat (squares) and rough (circles).
           (b) The height probability distribution for hydrophilic surface
           both flat (squares) and rough (circles). }
\end{figure}

Accordingly to the Wenzel equation, the apparent contact angle $\theta_0$
depends only on the surface roughness via the ratio $r=A/A_0$. 
Fig.~\ref{ca_ratio} shows that as $H$ decreases from $1$ to $0.4$ 
(i.e., $D_{\rm f}$ increases from 2 to 2.6), $A/A_0$ increases by $\sim 50\%$.
However, the MD-calculations show that the apparent contact angle $\theta_0$ 
is almost independent of the fractal dimension, see Fig.~\ref{ca_vs_H}.
This is also illustrated in Fig.~\ref{rms.H.ca.vs.A} which shows the contact angle
as a function of the (normalized) surface area $A/A_0$ for both cases. 
Thus the Wenzel equation cannot be used in the present situation. 
This is consistent with a visual inspection of the liquid-substrate
interface which shows that on the rough substrates,
the droplet is ``riding'' on the asperity tops of the substrate, i.e., the
droplet is in a Cassie-like state. In order to quantitatively verify this,
we have calculated the distances $u(x,y)$ between the bottom
surface of the liquid drop and the rough substrate surface in the (apparent) 
contact area. From the distribution\cite{p.u.MD.PRL}
$$P(u)=\langle \delta [u-u(x,y)]\rangle $$ 
of these 
distances [see Fig.~\ref{histogram_droplet}(a)] we obtain
the fraction $\psi$ of the (projected) surface area where contact occurs:
$$\psi = \int_0^{u_1} du \ P(u),$$
where $u_1$ is a cut-off distance to distinguish between contact
and no-contact regions, which has to be comparable to the typical
bond distance (we use $u_1=4$ \AA). Note that due to the thermal fluctuations 
$\psi=\psi_0$ for flat surface is less than $1$. Using the normalized 
$\phi = \psi/\psi_{0}$, the Cassie model (with $r=1$) predicts the variation of the contact 
angle with $\sigma$ and $H$ given in Fig.~\ref{ca_vs_rms} and \ref{ca_vs_H} 
(square points).

Fig.~\ref{ca_vs_rms} shows that the apparent contact angle $\theta_0$ 
increases strongly with increasing rms-roughness amplitude, at fixed
fractal dimension $D_{\rm f}=2.2$, while it is nearly independent of the 
fractal dimension $D_{\rm f}$ (see Fig.~\ref{ca_vs_H}).
Since increasing the fractal dimension at constant rms-roughness amplitude 
mainly increases the short-wavelength roughness, we conclude that the
nanoscale wavelength roughness doesn't matter so much in determining the 
contact angle for hydrophobic surfaces, while the long wavelength roughness 
plays an important role. We attribute this fact to the strong thermal 
fluctuations in the height (or width) $u$ of the liquid-solid interface which 
occur at the nanoscale even for the flat substrate surface.
Note also that in our model the wall-wall interaction
is long-ranged, decaying effectively as $\sim 1/u^3$, so there will be a 
contribution to the interfacial energy also for non-contacting surfaces
which, of course, is not rigorously included in the macroscopic Cassie model.

\subsection{Dynamic contact angle: Contact angle hysteresis}
\label{sec:dynamicangle}

The advancing contact angle $\theta_{\rm a}$ is measured when the solid/liquid
contact area increases, while the receding contact angle $\theta_{\rm r}$ is
measured when the contact area shrinks. If the difference $\theta_{\rm a}-\theta_{\rm r}$
is nonzero, the liquid-substrate system exhibits {\it contact angle hysteresis}.

\begin{figure}
  \begin{center}
  \includegraphics[width=0.45\textwidth]{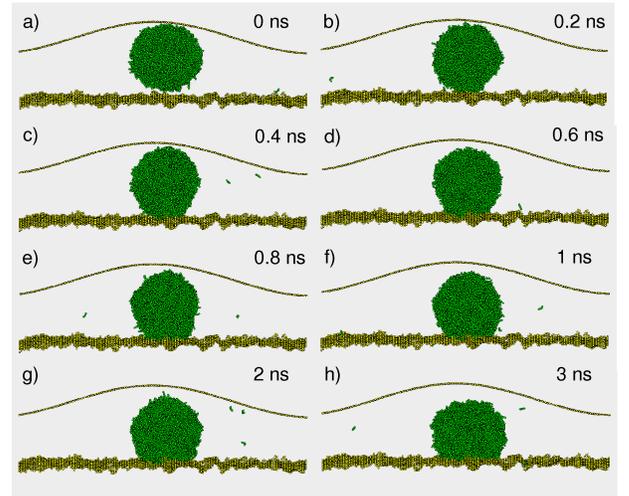}
  \end{center}
    \caption{ \label{advancing_theta}
           The advancing contact angle $\theta_{\rm a}$ evolution for
           hydrophobic nanodroplet. $\theta_{\rm a}$ is measured when the
           solid/liquid contact area increases.}
\end{figure}

\begin{figure}
  \begin{center}
  \includegraphics[width=0.45\textwidth]{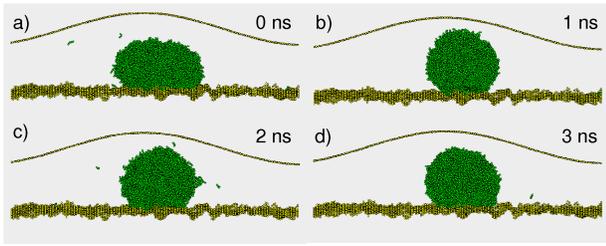}
  \end{center}
    \caption{ \label{receding_theta}
           The receding contact angle $\theta_{\rm r}$ evolution for
           hydrophobic nanodroplet. $\theta_{\rm r}$ is measured when the
           solid/liquid contact area shrinks.}
\end{figure}

\begin{figure}
  \begin{center}
  \includegraphics[width=0.45\textwidth]{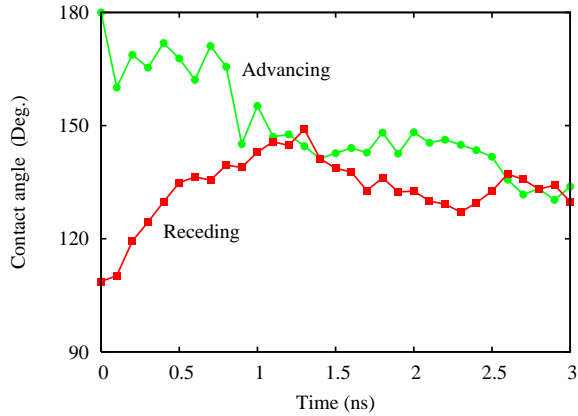}
  \end{center}
    \caption{ \label{hysteresis_ep4}
           The advancing (circles) and receding (squares) contact angle
           $\theta$, for hydrophobic substrate, as a function of time.
           The root-mean-square roughness of the substrate is ${\rm rms}=4.8$ \AA.
           $\epsilon=4 \ {\rm meV}$ and $r_{0}=3.28$ \AA. The thermal equilibrium
           contact angle has been reached after a few nanoseconds, irrespective of whether
           the initial contact angle is larger or smaller than the equilibrium
           angle.}
\end{figure}

\begin{figure}
  \begin{center}
  \includegraphics[width=0.4\textwidth]{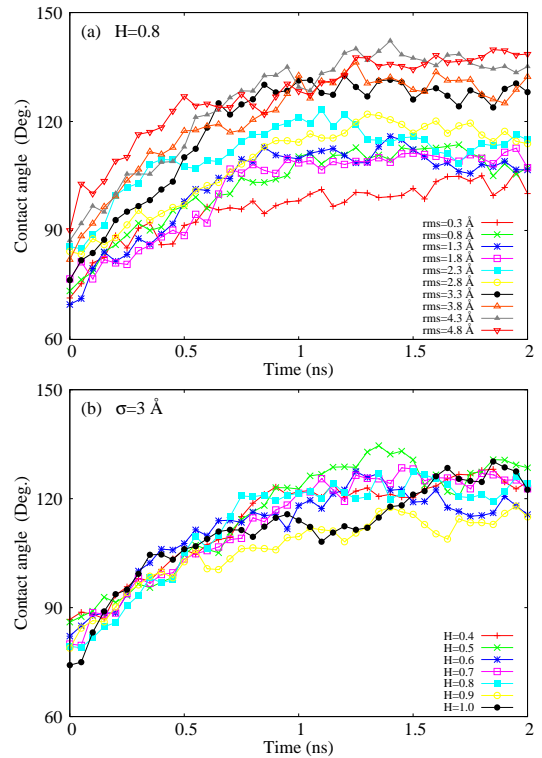}
  \end{center}
   \caption{\label{multi_pic_receding_Ugo.eps}
            The time evolution of the receding contact angle for the hydrophobic droplets,
            on the substrates with (a) various root-mean-square (rms) roughness while the 
            Hurst exponent $\rm H=0.8$, and (b) various Hurst exponent $\rm H$ while the 
            root-mean-square roughness (${\rm rms}=3$ \AA).}
\end{figure}

\subsubsection{Hydrophobic surfaces}
\label{sec:hydrophobic}

Figures \ref{advancing_theta} and \ref{receding_theta} show the time-evolution of the 
advancing contact angle $\theta_{\rm a}$ and
receding contact angle $\theta_{\rm r}$, respectively, for
a nanodroplet on a rough, hydrophobic substrate. 
The former has been obtained by placing the droplet close to the substrate, so that
the drop will spontaneously spread under the adhesive interaction with the substrate.
The contact angle evolves in time from $\theta=180^{\circ}$ (non-contact) in panel 
\ref{advancing_theta}(a) to its asymptotic value $\theta_{a}=140^{\circ}$, reached 
after $3 \rm ns$ in panel \ref{advancing_theta}(h).

The receding contact angle was simulated by squeezing the droplet into a pancake-like
shape with the upper wall. The interaction between the atoms of the upper wall and
drop atoms is given by the repulsive term of a Lennard-Jones potential, i.e.
$V(r)=4\epsilon_{0}(r_{0}/r)^{12}$. The lack of attraction with the top surface allows
us to suddenly pull the wall up, leaving the drop in the configuration of panel
\ref{receding_theta}(a). The free drop increases the contact angle up to the 
asymptotic value $\theta_{r}$. Figure \ref{hysteresis_ep4} shows the 
time evolution of the contact angle for these two cases. Note the strong time oscillations
of the contact angle which are due to
the rearrangement of the liquid molecules at
the solid-liquid interface. The corresponding energy barriers are small
compared to the thermal fluctuations (see
Sec.~IIC). However, after a few nanoseconds we find that both
the receding and advancing contact angle
fluctuate around the same average value; thus no contact angle hysteresis is
observed for the hydrophobic system.

To be sure that in any system there is no contact angle hysteresis, we performed extensive simulations
on various substrates with different
root-mean-square roughness amplitudes (rms) (see Fig.~\ref{multi_pic_receding_Ugo.eps}(a)). 
and with different
 Hurst exponents $H$ (see Fig.~\ref{multi_pic_receding_Ugo.eps}(b)),
The receding contact angle reaches its asymptotic value within about $2$ or $3$ nanoseconds.
In Figure \ref{multi_pic_receding_Ugo.eps} one can see a relatively broad
range of receding contact angles for substrates wih different rms roughness.
Conversely, substrates with different Hurst exponents show nearly
the same contact angle.
The root-mean-square roughness is mainly determined by
long wavelength roughness of the surface. 
Increasing the fractal dimension $D_{\rm f}$ signifies that the short wavelength roughness increases.
Thus, one can see that the contact angle 
is more sensitive to the long wavelength roughness of the substrate than to the
short wavelength roughness.
This agrees with the results in Fig.~\ref{ca_vs_rms} and ~\ref{ca_vs_H}.

A comparison of these results with those of the corresponding simulations for the advancing
contact angle, confirms that there is no hysteresis.
This is in drastic contrast to simulation studies we have performed 
for hydrophilic surfaces (see below), where surface roughness results in strong pinning 
of the boundary line; for such surfaces it is therefore
impossible to study static droplet contact angles (as observed on 
macroscopic time scales) using molecular dynamics. 

Comparing the form of $P(u)$ for the flat and the most rough surfaces shows 
that the system is in a Cassie-like state, but at the nanoscale the difference 
between the Cassie state and the Wenzel state is not so large due to the 
thermal fluctuations. That is, already for the flat surface strong thermal fluctuations
at the interface result in nanosized regions where the separation between the solid and the liquid
is much larger than the natural (low-temperature) binding separation. When
the substrate is rough the fluctuations become even larger and the system is in a state which
is more Cassie-like than Wenzel-like. 
This also explains why no hysteresis is observed: 
The Wenzel state is probably the (low temperature) 
energy minimum state (see Sec.~IID), but squeezing the droplet 
into a pancake shape does not push the system permanently 
into the Wenzel state (where pinning effects may be very important)
because even if it would go into this state temporarily, the free energy 
barrier separating the Cassie and Wenzel states is so small that
thermal fluctuations would quickly  kick it back to the Cassie-like state.

\begin{figure}
  \begin{center}
  \includegraphics[width=0.45\textwidth]{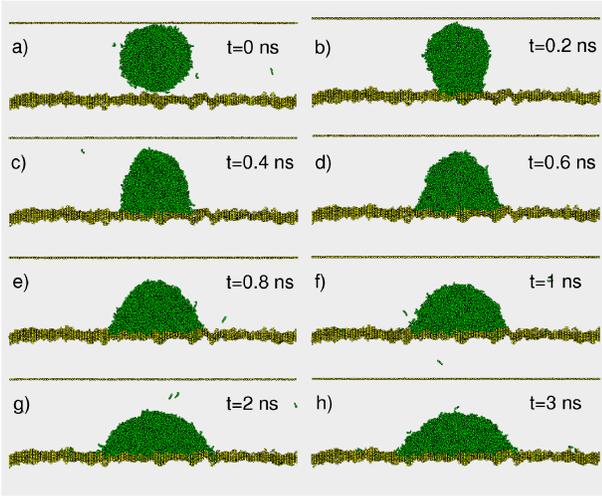}
  \end{center}
 \caption{ \label{adv_ep8_snapshots}
           Advancing contact angle evolution for hydrophilic nanodroplet.
           The root-mean-square roughness of the substrate is ${\rm rms}=4.8$ \AA.
           The energy parameter and the equilibrium distance in L.-J. potential
           are $\epsilon=8 \ {\rm meV}$ and $r_{0}=3.28$ \AA.}
\end{figure}

\begin{figure}
  \begin{center}
  \includegraphics[width=0.45\textwidth]{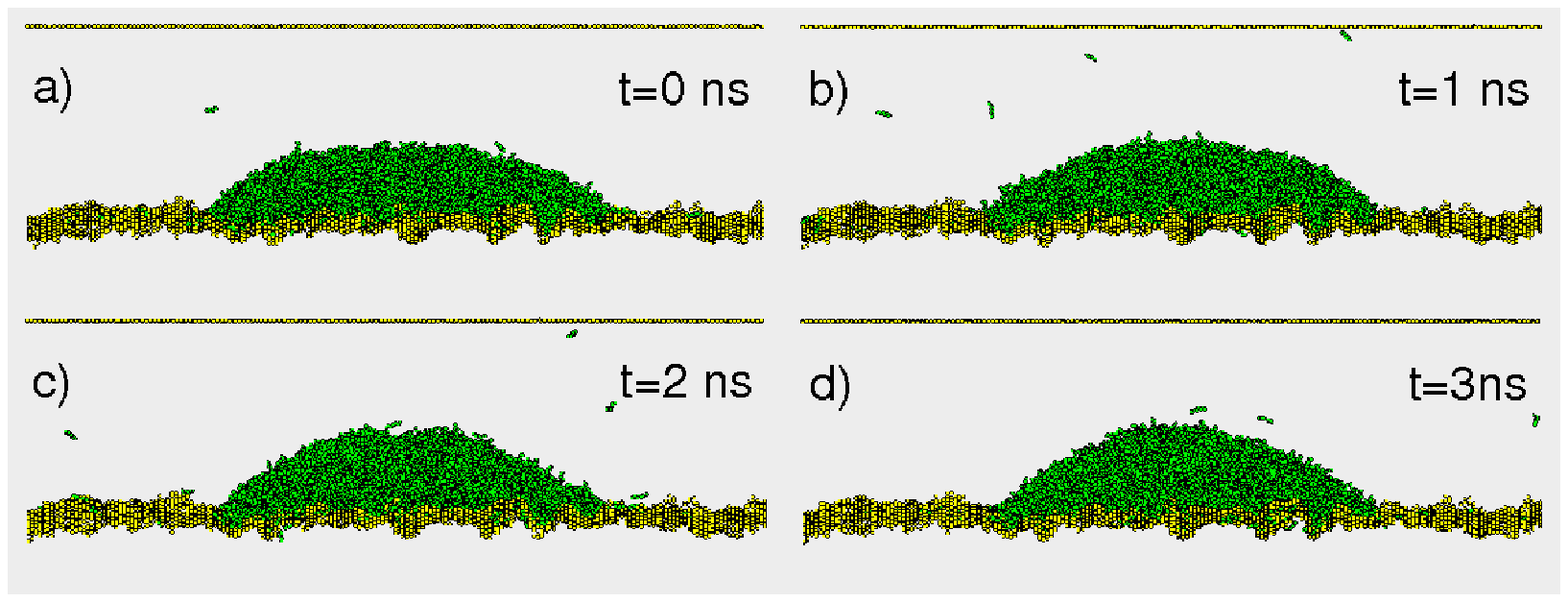}
  \end{center}
    \caption{ \label{rec_ep8_snapshots}
           Receding contact angle evolution for hydrophilic nanodroplet.
           The root-mean-square roughness of the substrate is ${\rm rms}=4.8$ \AA.
           The energy parameter and equilibrium parameter in L.-J. potential
           are $\epsilon=8$ meV and $r_{0}=3.28$ \AA.}
\end{figure}

\begin{figure}
  \begin{center}
  \includegraphics[width=0.45\textwidth]{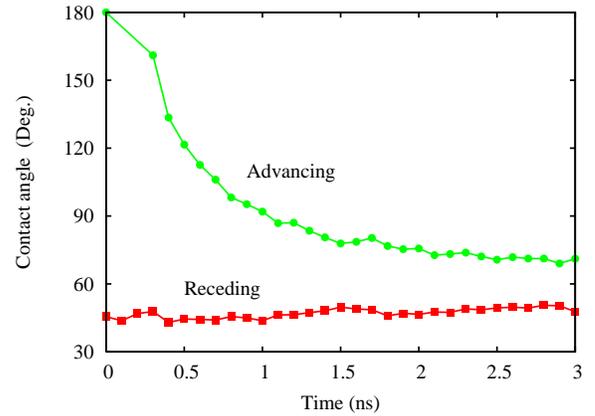}
  \end{center}
    \caption{ \label{hysteresis_ep8}
           The advancing (circles) and receding (squares) contact angle
           $\theta$, for the hydrophilic substrate, as a function of time.
           The root-mean-square roughness of the substrate is ${\rm rms}=3$ \AA,
           and the L.-J. substrate-liquid interaction parameters
           $\epsilon=8$ meV and $r_{0}=3.28$ \AA.}
\end{figure}

\subsubsection{Hydrophilic surfaces}
\label{sec:hydrophilic}

Let us now consider the case where the liquid droplet contact angle on the flat 
surface $\theta < 90^{\circ}$
(hydrophilic system). We choose the energy parameter and the 
equilibrium distance in L.-J. potential,
associated with the liquid-solid atom interaction, as 
$\epsilon=8 \ {\rm meV}$, $r_{0}=3.28$ \AA\ respectively. This gives $\theta \approx 70^{\circ}$. 
Figures \ref{adv_ep8_snapshots}, \ref{rec_ep8_snapshots} and \ref{hysteresis_ep8} for hydrophilic droplet,
are analogous to Figures \ref{advancing_theta}, \ref{receding_theta} and \ref{hysteresis_ep4} 
respectively, for hydrophobic droplet.
In Fig.~\ref{adv_ep8_snapshots} we
show the time dependence of the advancing contact angle for the hydrophilic nanodroplet.
The root-mean-square roughness of the substrate is ${\rm rms} =4.8$ \AA.

In Fig.~\ref{rec_ep8_snapshots} we
show the time evolution of the 
receding contact angle for the hydrophilic nanodroplet.
    
Fig.~\ref{hysteresis_ep8}
shows the advancing (circles) and receding (squares) contact angle
$\theta$ as a function of time, on a hydrophilic substrate with
root-mean-square roughness ${\rm rms}=3$ \AA.
Note that the thermal equilibrium
contact angle cannot be reached on the time-scale of the simulations.
Note also that the fluctuation in the contact angle are much smaller than for the hydrophobic system
(Fig.~\ref{hysteresis_ep4}). This is, of course, due to the fact that for the hydrophilic system
the liquid-substrate interaction is much stronger, and 
the barriers for the rearrangement of liquid molecules
at the substrate-liquid interface much higher than for the hydrophobic system.

Finally, in Fig.~\ref{histogram_droplet}(b) we show
the height probability distribution for the hydrophilic surface
for both the flat (squares) and rough (circles) hydrophilic surface. Note that the
fluctuations in the liquid-solid separation at the interface is much smaller
on the hydrophilic surface than on the hydrophobic surface [\ref{histogram_droplet}(a)].

\begin{figure}
  \begin{center}
  \includegraphics[width=0.45\textwidth]{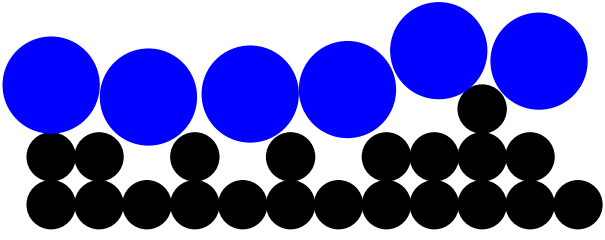}
  \end{center}
    \caption{ \label{cannotfollow}
The natural separation between the molecules in a liquid 
is usually considerably larger than the atom-atom separation on the substrate
surface. This implies that the fluid molecules cannot ``follow'' the atomic
scale roughness so that the fluid molecules will naturally
be in a Cassie-like state with respect to the shortest substrate roughness
components determined by the substrate nearest neighbor atom-atom separation.} 
\end{figure}

\section{Discussion}
\label{sec:discussion}

In most practical cases it is not possible to modify the surface roughness 
without simultaneously affecting the chemical nature of the surface. 
While this is obvious for crystalline materials, where surface
roughening will result in the exposure of new lattice planes with different 
intrinsic surface energy, it may also hold for amorphous-like materials,
where surface roughening may result in a more open atomic surface structure, 
with an increased fraction of (weak) unsaturated bonds.
In our model study a similar effect occurs, and some fraction of the change 
in the contact angle with increasing root-mean-square amplitude may be 
associated with this effect. However, the most important result of our study, 
namely that the contact angle is mainly determined by the long-wavelength 
roughness, should not be affected by this fact.

Another reason for why the short-wavelength (atomic) roughness may influence
the liquid contact state differently from long-wavelength roughness has to do
with the fact that the natural separation between the molecules in a liquid 
is usually considerably larger than the atom-atom separation on the substrate
surface. This implies that the fluid molecules cannot ``follow'' the atomic
scale roughness (see Fig.~\ref{cannotfollow}) so that the fluid molecules will naturally
be in a Cassie-like state with respect to the shortest substrate roughness wavelength
components, determined by the substrate nearest neighbor atom-atom separation. 

It is important to note that the discussion in this paper is also relevant for 
the contact between macroscopic liquid drops and rough substrates. That is, if
the solid-liquid interface is studied with nanometer resolution $\lambda$, then the 
strong fluctuations (in time and space) at the liquid-solid 
interface discussed above will also be observed for a macroscopic droplet,
and the interfacial energy $\gamma_{\rm sl}(\zeta)$ for $\zeta = D/\lambda$ 
(where $D$ is the diameter of the nominal contact area between the (macroscopic)
liquid droplet and the substrate) will be the same as obtained above for 
nanodroplets at the same resolution $\lambda$.

\section{Summary and conclusion}
\label{sec:conclusions}

We have discussed under which condition the
Wenzel or
Cassie state is favorable on randomly rough surfaces.
We performed molecular dynamics simulations to study contact angle and the contact 
angle hysteresis on hydrophobic and hydrophilic surfaces. 
The contact angle on hydrophobic surfaces depends strongly on the root-mean-square 
roughness of the substrate, but is nearly independent of 
the fractal dimension.
For hydrophobic surfaces, there is no contact angle hysteresis due to 
strong thermal fluctuations at the nanoscale. For hydrophilic surfaces we 
observe contact angle hysteresis due to pinning effects resulting from the much higher energy
barriers for rearrangement of liquid molecules at the solid-liquid interface.
This indicates that on randomly rough
hydrophobic surfaces the Cassie-like state often prevails, at least 
for nanoscale droplets.
We have found that thermal fluctuations play an important role at the nanoscale, which 
leads to the enhanced hydrophobicity by surface roughness. It is of particular importance
to design and build superhydrophobic surfaces. 
 
\section{Appendix: Distribution of surface slopes for randomly rough surfaces}

In this appendix we derive the distribution of surface slopes for randomly rough
surfaces. We discuss under which conditions one expects the Wenzel and Cassie states
to prevail.

\vskip 0.3cm
{\bf The surface area $A$ and the average surface slope $\xi$}

Consider a randomly rough surface and let $h({\bf x})$ denote the
height profile measured from the average plane 
so that $\langle h({\bf x})\rangle = 0$, where $\langle .. \rangle$ stands for ensemble averaging,
or (equivalently) averaging over the surface area. 
We assume that $h({\bf x})$ is a Gaussian random variable
characterized by the power spectrum
$$C(q) = {1\over (2 \pi )^2} \int d^2 x \ \langle 
h({\bf x})h({\bf 0})\rangle e^{-i{\bf q}\cdot {\bf x}}.$$
Note that if we write
$$h({\bf x}) = \int d^2 q \ h({\bf q}) e^{i {\bf q}\cdot {\bf x}}$$
where
$$h({\bf q}) = {1\over (2 \pi)^2} \int d^2 x \ h({\bf x}) e^{-i {\bf q}\cdot {\bf x}}$$
then
$$\langle h({\bf q}) h({\bf q}')\rangle = \delta ({\bf q}+{\bf q}') C(q).\eqno(A1)$$
Sometimes it is also convenient to use
$$\langle h({\bf q}) h(-{\bf q})\rangle 
= {A_0\over (2\pi)^2} C(q),\eqno(A2)$$
where $A_0$ is the surface area. In deriving (A2) we have used
that
$$\delta ({\bf q}- {\bf q}) = {1\over (2 \pi)^2} \int d^2x \ e^{i({\bf q}- {\bf q})\cdot {\bf x}} 
={A_0 \over (2\pi )^2}.$$
 
If the surface roughness amplitudes $h({\bf q})$ are assumed to be Gaussian random variables,
one can show that the (normalized) surface area\cite{P.T}  
$$r={A\over A_0} = \int_0^\infty dx \ \left (1+x\xi^2\right )^{1/2}e^{-x}$$
where
$$\xi^2 = \int d^2q \ q^2C(q) = 2 \pi 
\int_0^\infty dq \ q^3C(q) \,.
$$
Let us calculate the rms surface slope. We get
$$\langle (\nabla h)^2 \rangle = \int d^2q d^2q' \ (iq_\alpha)(iq'_\alpha) \langle h({\bf q})
h({\bf q}')\rangle e^{i({\bf q}+{\bf q}')\cdot {\bf x}}$$
Using (A1) this gives
$$\langle (\nabla h)^2 \rangle = \int d^2q \ q^2 C(q) = \xi^2$$
Thus, {\it for a Gaussian random surface both the average slope and the increase in the surface area is
determined by the parameter $\xi$}. For non-random surfaces this is no longer the case.

\vskip 0.3cm
{\bf Surface slope probability distribution}

Let $h({\bf x},\zeta )$ denote the height profile after having smoothed out surface roughness
with wavelength shorter than $\lambda = L/\zeta$. For example, 
we may define
$$h({\bf x},\zeta) = \int_{q<q_1} d^2 q \ h({\bf q}) e^{i {\bf q}\cdot {\bf x}},$$
where $q_1 = q_L \zeta$ (where $q_L=2\pi /L$). We will refer to $\zeta$ as the magnification. 
Thus, when we study the surface at the magnification $\zeta$ we will only detect surface roughness
with wavelength components larger than $\lambda = L/\zeta$. 

We will now derive an equation of motion for the surface slope probability distribution function
$$P({\bf s},\zeta ) = \langle \delta ({\bf s} - \nabla h({\bf x},\zeta))\rangle. $$
We assume that the surface roughness amplitudes $h({\bf q})$ are independent random variables.
In this case,
if we write
$$h({\bf x},\zeta+\delta \zeta) = h({\bf x},\zeta)+\delta h,$$ 
we get
$$P({\bf s},\zeta +\delta \zeta) = 
\langle \delta ({\bf s} - \nabla h({\bf x},\zeta+\delta \zeta))\rangle $$
$$=\int d^2s' \langle \delta ({\bf s'} - \nabla \delta h)\rangle
\langle \delta ({\bf s}-{\bf s'} - \nabla h ({\bf x},\zeta))\rangle$$
$$=\int d^2s' \langle \delta ({\bf s'} - \nabla \delta h)\rangle
P({\bf s}-{\bf s'},\zeta).\eqno(A3)$$
But
$$\langle \delta ({\bf s'} - \nabla \delta h)\rangle = {1\over (2\pi )^2}\int d^2q \langle
e^{i{\bf q}\cdot ({\bf s'}-\nabla \delta h)}\rangle$$
$$= {1\over (2\pi )^2}\int d^2q \left (1- {1\over 2}\langle
({\bf q}\cdot \nabla \delta h)^2 \rangle \right )
e^{i{\bf q}\cdot {\bf s'}}$$
$$=\delta ({\bf s}')+{1\over 2} \langle \left (\nabla_\alpha \delta h \right )
\left (\nabla_\beta \delta h \right )\rangle {\partial \over \partial s'_\alpha}
{\partial \over \partial s'_\beta} \delta ({\bf s}').$$
Substituting this result into (A3) and expanding the left-hand-side to linear order in $\delta \zeta$ gives
$$ 
{\partial P \over \partial \zeta} ({\bf s},\zeta ) 
= 
{1\over 2\delta \zeta } \langle \left (\nabla_\alpha \delta h \right )
\left (\nabla_\beta \delta h \right )\rangle
{\partial \over \partial s_\alpha}
{\partial \over \partial s_\beta} P({\bf s},\zeta).\eqno(A4)$$
But
$$\langle \left (\nabla_\alpha \delta h \right )
\left (\nabla_\beta \delta h \right )\rangle
 = \int_{q_L\zeta}^{q_L(\zeta+\delta \zeta)} d^2q d^2q'$$
$$\times (iq_\alpha )(iq'_\beta ) \langle h({\bf q}) h({\bf q'}) \rangle e^{({\bf q}+{\bf q}')\cdot {\bf x}}$$
$$=
\int_{q_L\zeta}^{q_L(\zeta+\delta \zeta)} d^2q \
q_\alpha q_\beta C(q)$$
$$=
{1\over 2} \delta_{\alpha \beta} \int_{q_L\zeta}^{q_L(\zeta+\delta \zeta)} d^2q \ 
q^2 C(q)$$
$$=
\pi \delta_{\alpha \beta} q_L \delta \zeta q^3 C(q).$$
Thus
$${1\over 2 \delta \zeta} \langle \left (\nabla_\alpha \delta h \right )
\left (\nabla_\beta \delta h \right )\rangle = 
{\pi \over 2} \delta_{\alpha \beta} 
q_L q^3 C(q)$$
Substituting this result in (A4) gives the following diffusion-like equation for $P({\bf s},\zeta)$:
$$ 
{\partial P \over \partial \zeta} 
= 
D(\zeta ) \nabla_s^2 P\eqno(A5)$$
where
$$\nabla_s^2 = {\partial \over \partial s_\alpha}
{\partial \over \partial s_\alpha}$$
where the ``diffusivity''
$$D(\zeta) = {\pi \over 2} q_L q^3 C(q),\eqno(A6)$$
with $q=q_L \zeta$.

\vskip 0.3cm

{\bf Solution of the diffusion equation}

The function $P({\bf s},\zeta)$ describes the probability to observe a surface slope
or gradient ${\bf s} = \nabla h({\bf x})$ when the system is studied at the magnification $\zeta$.
When the system is studied at the lowest magnification $\zeta = 1$ the surface appears 
flat and smooth so that the gradient vanishes, i.e.,
$$P({\bf s},1) = \langle \delta ({\bf s}-\nabla h({\bf x},1))\rangle = \delta ({\bf s}).\eqno(A7)$$
We also require that there is no infinite high slopes, i.e.,
$$P({\bf s},\zeta) \rightarrow 0 \ \ \ {\rm as} \ \ \ |{\bf s}| \rightarrow \infty. \eqno(A8)$$
Let us determine the solution to (A5) 
which obeys the ``initial'' condition (A7) and the boundary condition (A8).
It is clear that the solution is given by
$$P({\bf s},\zeta) = {1\over \pi s_1^2} e^{-(s/s_1)^2}, \eqno(A9)$$
where the width parameter $s_1(\zeta)$ depends on the magnification:
$$s_1^2 = 4 \int_1^\zeta d\zeta' D(\zeta') 
= {2 \pi } \int_{q_L}^{\zeta q_L} dq q^3 C(q) = \xi^2 (\zeta). $$  
Note that $P$ is normalized,
$$\int d^2s \ P({\bf s}, \zeta) = 1,$$
and that the width of the Gaussian distribution $P$ increases 
with increasing resolution, i.e., when the
surface is studied at higher and higher resolution, 
steeper and steeper surface slopes will be detected.

\vskip 0.3cm

{\bf Distribution function $P(s_0,\zeta)$}

In what follows we will need the fraction $P(s_0,\zeta)$ of the total surface area
where the slope $s < s_0$, where $s_0$ is a fixed number between
zero and infinite. Let $A_\perp (\zeta)$
be the surface area, projected on the $xy$-plane, 
where the surface slope $s < s_0$. We then have
$$P(s_0,\zeta ) = A_\perp (\zeta) /A_0.$$
Using the definition
$$P({\bf s},\zeta) = \langle \delta ({\bf s}-\nabla h({\bf x},\zeta))\rangle$$
$$= {1\over A_0} \int d^2x \ 
\delta ({\bf s}-\nabla h({\bf x},\zeta))$$
we get
$$P(s_0,\zeta) = \int_{s < s_0}d^2s \ P({\bf s},\zeta).$$
Using (A9) this gives 
$$P(s_0,\zeta ) = 1-e^{-(s_0/s_1)^2}.\eqno(A10)$$

\begin{figure}
\includegraphics[width=0.3\textwidth]{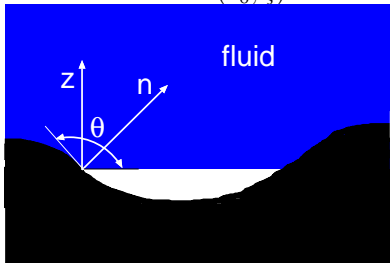}
\caption{\label{Contact} 
A fluid in contact with a rough substrate. The contact angle $\theta$
is determined by Young's equation.} 
\end{figure}

\vskip 0.3cm

{\bf Surface area with slope below ${\rm tan}\theta$}

Consider a liquid in contact with a rough substrate. The contact angle $\theta$
is determined by Young's equation:
$${\rm cos} \theta = {\gamma_{\rm sv}-\gamma_{\rm sl}\over \gamma_{\rm lv}}$$
where $\gamma_{\rm sl}$, $\gamma_{\rm sv}$ and $\gamma_{\rm lv}$ 
are the solid-liquid, solid-vapor and liquid-vapor
interfacial energies, respectively. Note that if ${\bf n}$ is the normal to the solid surface
and ${\bf z}$ the normal to the liquid surface, which we assume to be parallel to the
average surface plane (see Fig.~\ref{Contact}), then
${\rm cos} \theta = - {\bf z}\cdot {\bf n}$. Since
$${\bf n} = {(-\nabla h,1)\over \left(1+(\nabla h)^2\right )^{1/2} }$$
we get
$${\rm cos} \theta = - \left (1+(\nabla h )^2\right )^{-1/2}, \ \ \ \ 
\ \ \ |{\rm tan \theta}| = |\nabla h |. $$
Thus, using (A10) the fraction of the surface where the surface slope is below ${\rm tan} \theta$
is
$$P({\rm tan} \theta ) = 1-e^{-({\rm tan} \theta/\xi)^2}. \eqno(A11)$$


\begin{thebibliography}{999}
\bibitem{cappilary}
The surface energy of a liquid does, in fact, also depend on the magnification
because of thermally excited capillary waves which contribute to the surface energy of
a liquid. However, the dependence of $\gamma_{\rm lv}(\zeta)$
on the magnification $\zeta$ is rather weak and we will neglect this effect in the present
study.
\bibitem{comment1}
Here we assume that the surface free energy per unit area does not depend on the
orientation of the substrate surface, which may be a reasonable approximation for many 
amorphous materials, but which in general fails for crystalline materials.
\end{thebibliography}
\end{document}